\documentclass{Rinton-P10x7}

\usepackage{epsf}

\begin{document}

\title{DARK MATTER EQUATION OF MOTION AND DENSITY PROFILES}

\author{ANTONIO R. MONDRAGON AND ROLAND E. ALLEN}

\address{Center for Theoretical Physics, Texas A\&M University, \\[0 pt]
College Station, Texas 77843, USA}

\maketitle

\abstracts{Cold dark matter simulations appear to disagree with the
observations on small distance scales. Here we consider a modified
description of  CDM particles which is implied by a new fundamental theory:
In this picture, the dark matter is composed of supersymmetric WIMPs, but
they are scalar bosons with an unconventional equation of motion. The
modified dynamics leads to much weaker gravitational binding, and therefore
to a reduced tendency to form small clumps and central cusps.}

Cold dark matter simulations provide a satisfactory description 
of the evolution of large-scale structure in the universe, 
but appear to disagree in two respects with the observations 
relevant to galactic halos$^{1-13}$: First, the simulations 
demonstrate that standard CDM has too strong a 
tendency to form relatively small clumps. 
Second, the simulations yield cusps in the CDM density 
$\rho \left( r \right)$ of the form\cite{navarro96}
\begin{equation}
\rho \left( r \right) \propto \frac{1}{\left( r/r_{s} \right)
\left(1+  r/r_{s} \right)^{2}} \propto r^{-1} \; \mbox{as} 
\; r \rightarrow 0
\end{equation}
whereas analyses of the observational data seem to indicate a 
flattening of the density as $r \rightarrow 0$.  

In this paper we consider a modified description of dark matter 
particles which is implied by a new fundamental theory\cite{allen1,allen2}:
The dark matter is composed of supersymmetric WIMPs, just as in more 
conventional models, but in the present picture these 
are scalar bosons with an unconventional equation of motion (rather 
than, e.g., neutralinos with standard relativistic or nonrelativistic 
dynamics). The modified dynamics leads to much weaker binding 
in a gravitational field, and therefore to a reduced tendency to form 
clumps and cusps. The compatibility of this modified dynamics with 
standard physics is discussed elsewhere\cite{allen1,allen2}. The 
modifications are significant only for (i) fermions at extremely high energy 
and (ii) fundamental scalar bosons that have not yet been observed.  
The modified dynamics retains many of the features of Lorentz 
invariance, such as rotational invariance, CPT invariance, and the 
requirement that $\omega = \left| \vec{p}\right|$ 
for massless particles. It also appears to be consistent with even the most 
sensitive experimental tests of Lorentz invariance.

To be more specific, in the present theory the dark matter is composed of 
fundamental scalar bosons with an R-parity of -1 and with the equation of motion
\begin{equation}
\eta ^{\mu \nu }\,\partial _{\mu }\partial _{\nu }\phi +
i\overline{m}\sigma ^{\mu
}\partial _{\mu }\phi -m^{2}\phi =0
\end{equation}
where $\eta ^{\mu \nu }=diag(-1,1,1,1)$ is the Minkowski metric tensor, 
$m$ is the particle mass, and $\overline{m}$ is an energy which 
is comparable to or larger than $1$ TeV. For a plane-wave solution
$\phi \propto \exp \left( i\,\vec{p}\cdot \vec{x}-i\,\omega 
\,t\right)$ this becomes
\begin{equation}
\left[ \left( \omega ^{2}-\vec{p}\,^{2}\right) +\overline{m}\left( \omega -
\vec{\sigma}\cdot \vec{p}\right) -m^{2}\right] \phi =0
\end{equation}
with the solutions
\begin{equation}
\omega =-\frac{1}{2}\overline{m}\pm \left[ \left( p\pm \frac{1}{2}\overline{m}\right)
^{2}+m^{2}\right] ^{1/2}\quad ,\quad p=\left| \vec{p}\right| .
\end{equation}
The $\pm $ signs are independent, but negative frequencies correpond to
antiparticles in the usual way\cite{allen2}. For simplicity, 
therefore, let us focus on the positive-frequency solution with particle 
velocity
\begin{equation}
v=\frac{\partial \omega }{\partial p}=\left[ 1+m^{2}\left( p+\frac{1}{2}
\overline{m}\right) ^{-2}\right] ^{-1/2}.
\end{equation}
The kinetic energy is
\begin{equation}
T=\int v\left( p\right) dp = m\left( \gamma -\gamma _{0}\right) 
\end{equation}
where
\begin{equation}
\gamma =\frac{1}{\sqrt{1-v^{2}}}\quad ,\quad \gamma _{0}=\frac{1}{\sqrt
{1-v_{0}^{2}}}\quad ,\quad v_{0}=\left[ 1+\left( \frac{2m}{\overline{m}}\right)
^{2}\right] ^{-1/2}.
\end{equation}
(As $\overline{m}\rightarrow 0$, we regain the standard result of special
relativity in units with $\hbar =c=1$: $T=m\left( \gamma - 1\right)$.) 
With an integration by parts, (6) can also be written in the form
\begin{equation}
T=pv+\left( \overline{m}/2 \right) \left( v-v_{0}\right) 
+m\left( \gamma ^{-1}-\gamma_{0}^{-1}\right). 
\end{equation}

Now consider a circular orbit of radius $r$ about a mass $M$. 
The general formula for the centripetal force implies that
\begin{equation}
pv/r=GMm/r^{2}
\end{equation}
with $V=-GMm/r$, so
\begin{equation}
E=T+V=\left( \overline{m}/2 \right) \left( v-v_{0}\right) 
+m\left( \gamma ^{-1}-\gamma_{0}^{-1}\right) .
\end{equation}
As $m/\overline{m}\rightarrow 0$, we obtain $v,v_{0}\rightarrow 1$ and $\gamma
^{-1},\gamma _{0}^{-1}\rightarrow 0$ (since $v_{0}\leq v\leq 1$). It follows
that 
\begin{equation}
E\rightarrow 0\quad \mbox{as}\quad m/\overline{m}\rightarrow 0
\end{equation}
and the particles are only very weakly bound for $m/\overline{m}\ll 1$. This is 
qualitatively the same result as in special relativity when $m \rightarrow 0$.

\begin{figure}[ht]
\centering
\epsfxsize=29pc \epsffile{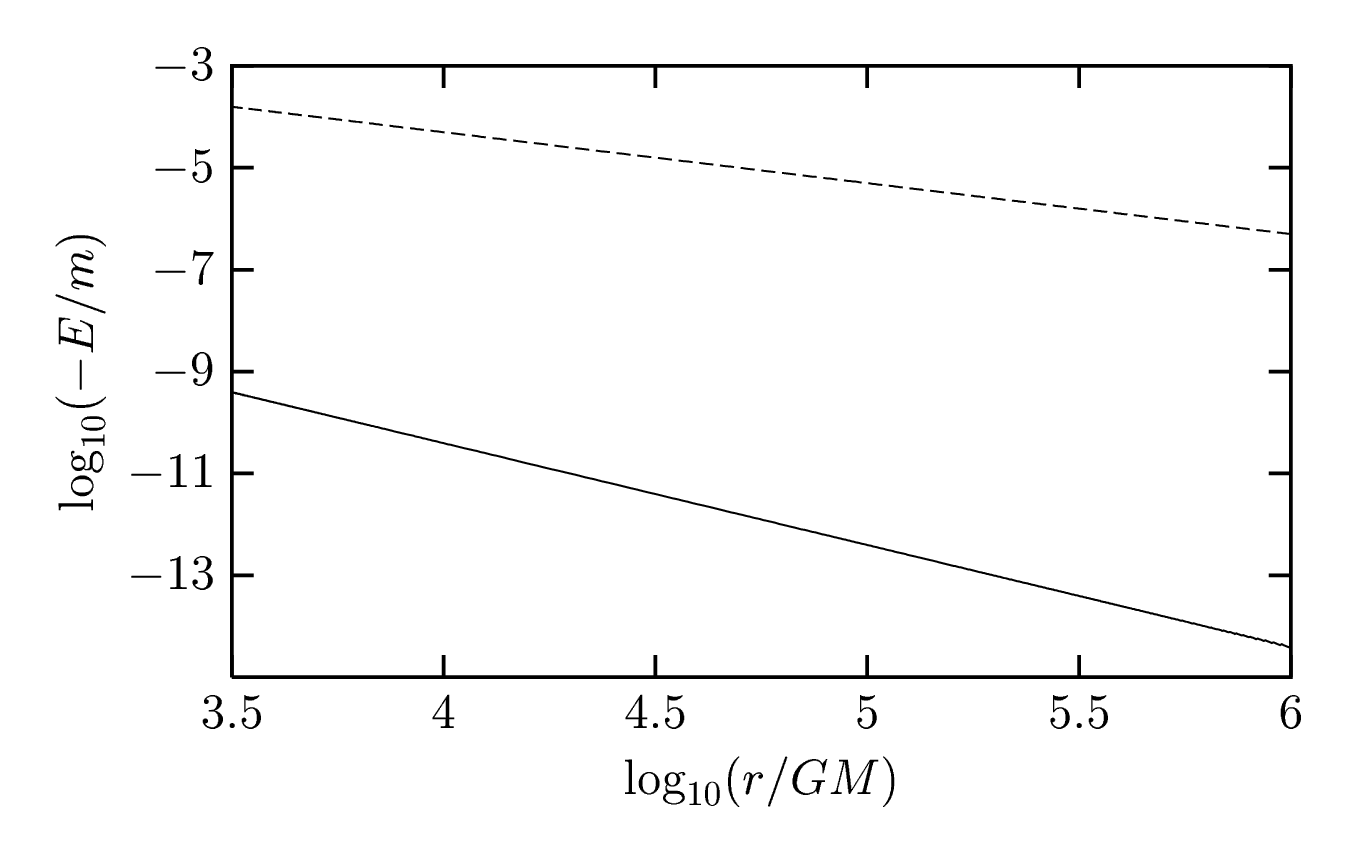}
\caption{Binding energy $-E \left( r \right)$, with the equation 
of motion (2) and $\overline{m}/m$ taken to be $10$ (solid line) compared 
with the dynamics of standard special relativity (dashed line). The 
binding is much weaker in the present theory.}
\end{figure}

Figure 1 shows the binding energy $-E$ as a function of $r$ with 
$\overline{m}/m$ taken to be $10$. In the same figure we show 
$-E \left( r \right)$ with the particles obeying the standard dynamics 
of special relativity. (Newtonian gravity was 
used, with the gravitational mass taken to be the rest mass. 
A graph for nonrelativistic dynamics would be indistinguishable from 
the dashed line of Fig. 1, since relativistic corrections are not 
important for $r \gg 2GM$.) The binding is orders 
of magnitude weaker in the present theory.

The slope in Fig. 1 for the standard dynamics is -1, since
\begin{equation}
-E\left( r\right) =-\left( T+V \right) = -\frac{1}{2} V = \frac{1}{2} 
\frac{GM}{r} m \propto \frac{1}{r} \quad \mbox{for} \quad r \gg 2GM.
\end{equation}
The slope for the dynamics of the present theory is -2, 
indicating that $-E\left( r\right) \propto 1/r^{2}$. One can, 
in fact, obtain a simple expression for $E\left( r\right)$, 
by expanding (5) in powers of $p$, substituting the resulting 
expression into (9), inverting to find $p$ as a function of $1/r$, 
and then substituting into (10). The final relation is
\begin{equation}
-E\left( r\right) =\frac{1}{2}\left( \frac{2m}{\overline{m}}\right)
^{3}\left( \frac{GM}{r}\right) ^{2} m \; \propto \frac{1}{r^{2}} 
\quad \mbox{for} \quad r \gg 2GM.
\end{equation}

Let us now consider a simplistic model in which 
\begin{equation}
\rho \left( r\right) =\frac{a}{r^{\alpha }}\quad \mbox{so that}\quad M\left(
r\right) =\frac{4\pi a}{3-\alpha }r^{3-\alpha }\quad \mbox{and}\quad F\left(
r\right) =-\frac{dV\left( r\right) }{dr}=-\frac{GM\left( r\right) m}{r^{2}}.
\end{equation}
The virial theorem states that 
\begin{equation}
\left\langle \vec{p}\cdot \vec{v}\right\rangle =-\left\langle \vec{F}\cdot 
\vec{r}\right\rangle=\left\langle GM\left( r\right) m /r\right\rangle .
\end{equation}
On the other hand, the equipartition theorem implies that 
\begin{equation}
\left\langle \vec{p}\cdot \vec{v}\right\rangle = 
\left\langle p^{k}\partial H / \partial p^{k}\right\rangle = 3\tau 
\end{equation}
where $H$ is the classical Hamiltonian and $\tau $ is the temperature in
units with $k_{B}=1$. Then an isothermal model corresponds to 
$\left\langle \vec{p}\cdot \vec{v}\right\rangle =$ constant, or
\begin{equation}
M\left( r\right) \propto r\quad , \quad V\left( r\right) \propto
\log \; r+ \; \mbox{constant}\quad , \quad 
\rho \left( r\right) \propto r^{-2} .
\end{equation}
This is, however, exactly the same qualitative result that we would have
obtained with standard nonrelativistic or relativistic dynamics.
Furthermore, it is easy to see that a complete solution for an isothermal
model with a Boltzmann distribution\cite{tremaine,chandra} also 
gives (17) at large $r$, which is again the same result as 
one obtains with standard dynamics.

The results of (11) or (13) and Fig. 1 lead us to conclude that the present
theory predicts much weaker gravitational binding than conventional cold
dark matter models. As a result, there is a greatly reduced tendency for
particles to be bound at small values of $r$, and one does not expect the
formation of cusps or clumps on small distance scales. On the other hand, it
is plausible that the successes of standard CDM models on large distance
scales may be preserved in the present theory, since an isothermal model
yields (17) for either standard dynamics or the modified dynamics
represented by (5). We intend to test these conclusions with simulations.

The present theory has implications for terrestrial dark matter 
searches\cite{cline}: Since the velocity $v$ in (9) or (15) 
is $\sim c$ rather than $10^{-3} c$, the momentum $p$ is reduced by a 
factor of roughly $10^{-3}$ compared to the expectation for 
conventional cold dark matter.

\section*{Acknowledgement}

This work was supported by the Robert A. Welch Foundation.

\end{document}